\documentclass[aps,prb,twocolumn,groupedaddress,showpacs]{revtex4}
\usepackage{epsfig}
\usepackage{graphicx}
\newcommand{\la}{\langle}
\newcommand{\ra}{\rangle}
\newcommand{\up}{\uparrow}
\newcommand{\dn}{\downarrow}

\begin{document}

\title{Transmission of correlated electrons through
sharp domain
walls in magnetic nanowires: a renormalization group approach}

\author{M. A. N. Ara\'ujo$^{a,b}$\footnote{On leave from Departamento de F\'{\i}sica,
Universidade de \'Evora, P-7000-671, \'Evora, Portugal}, V. K.
Dugaev$^{b,c,d}$, V. R. Vieira$^{b}$, J. Berakdar$^{d}$, and J.
Barna\'s$^{e,f}$} \affiliation{$^{a}$Department of Physics,
Massachusetts Institute of Technology, Cambridge MA 02139, U.S.A.}
\affiliation{$^{b}$CFIF and Departamento de F\'isica, Instituto
Superior T\'ecnico, Av. Rovisco Pais, 1049-001 Lisbon, Portugal}
\affiliation{$^c$Frantsevich Institute for Problems of Materials
Science, National Academy of Sciences of Ukraine, Vilde 5, 58001
Chernovtsy, Ukraine} \affiliation{$^d$Max-Planck Institut f\"ur
Mikrostrukturphysik, Weinberg 2, 06120 Halle, Germany}
\affiliation{$^e$Department of Physics, Adam Mickiewicz
University, Umultowska 85, 61-614 Pozna\'n, Poland}
\affiliation{$^f$Institute of Molecular Physics, Polish Academy of
Sciences, Smoluchowskiego 17, 60-179 Pozna\'n, Poland}

\date{\today }

\begin{abstract}
The transmission  of correlated electrons through  a domain wall in a
ferromagnetic one dimensional system is studied theoretically in
the limit of a domain wall width smaller or comparable to the
electron Fermi wavelength.  The domain wall gives rise to both
potential and spin dependent scattering of the charge carriers.
Using a poor man's renormalization group approach for the
electron-electron interactions, we obtain the low temperature
behavior of the reflection and transmission coefficients. The
results show that the low-temperature conductance is governed by
the electron correlations, which may suppress charge transport
without suppressing spin current. The results may account for a
huge magnetoresistance associated with a domain wall in ballistic
nanocontacs.
\end{abstract}

\maketitle
\section{Introduction}

Domain walls (DWs), i.e. the  boundaries separating different
domains of homogeneous magnetizations \cite{1} are recently a
subject of extensive theoretical and experimental investigations.
This renewed interest in DWs is stimulated by their possible
applications in magnetic logic elements and other nanoelectronics
and spintronics devices. Two effects associated with  DWs are of
particular interest. The first one is the way a DW affects
electronic transport, i.e., the associated magnetoresistance. The
crucial point here is that the influence of a single DW on the
resistance can be controlled by an external magnetic field
\cite{katine00,hong_nature_2004}. The second effect concerns the
influence of electric current on the DW behavior (DW motion,
magnetic switching)\cite{yamaguchi04}, which allows controlling of
DWs by means of an electric field.

Recent advances in experimental techniques have made possible the
determination of the resistance of a single DW in submicron
structured samples\cite{hong_nature_2004,9,10,11,12,13,14,15}. The
results on the single domain resistance are different in
magnitude and sometimes differ also in the sign. In the case
where the DW width is large on the scale set by the Fermi wavelength,
$\lambda_F$, of the carriers, the theory of the DW
contribution to electrical resistance is well-established
\cite{3,4,5,6,7,8}. The spin of the electron moving across the wall changes
its orientation quasi-adiabatically (or even adiabatically for very
thick DWs). However, the DWs formed at nanoconstrictions may be
atomically sharp\cite{d2,d3,d4,d5} and the spin  of an
electron crossing the wall does not change quasi-adiabatically.
 Accordingly, a
completely different approach to the transport theory through DWs
is required. This is particularly true for ferromagnetic
semiconductors which are considered to be most promising for
spintronic applications\cite{18}. Indeed, recent experiments on
magnetic nanostructures and  nanowires indicate that the presence
of DWs may result in a magnetoresistance (MR) as large as  several
hundreds \cite{12,garcia99} or even thousands
\cite{chopra02,ruster03} of percents, as opposed to the case of
thick on scale of $\lambda_F$  (or adiabatic) DWs in bulk metallic
ferromagnets. In the ballistic regime,
 the theoretical treatments towards explaining this effect
\cite{hoof99,kudrnovsky00,tagirov01,yavorsky02,dugaev03,zhuravlev03}
rely on the assumption that the DW is sharp enough to be treated
as a spin dependent scatterer for the charge carriers. The success
of these theories in explaining the extraordinary large MR is
moderate, in particular for  metallic ferromagnets such as  Ni
where some features of the physics governing the behaviour of MR
are still unclear\cite{simanek01}.

Another feature of the DWs created at nanoconstrictions is their
small lateral size (cross-section of the constriction). This small
size limits the number of quantum channels active in transport to
a few ones or even to a single one. Consequently, the constriction
behaves as a one- or quasi-one-dimensional system. In such a case,
the role of electron-electron interactions may be crucial
\cite{gogolin} for understanding the behavior and basic transport
characteristics of the DWs formed at nanoconstrictions. It is
well-established that  electronic correlations in a one-channel
wire result in a non-Fermi-liquid behavior - thus  forming a
Luttinger liquid \cite{tomonaga,luttinger}. It is also known that
an impurity present in the 1D Luttinger liquid suppresses the
linear conductivity, which vanishes even for a weak impurity
scattering potential\cite{kane1,kane2}. This can be traced back to
a vanishing density of states at the Fermi level. At finite
applied voltages the transport through the wire does not vanish
due to the nonlinearity of the current-voltage
characteristics\cite{kane1}. Since a sharp domain wall acts in a
one-channel wire as a localized spin-dependent scattering center,
one can expect a strong influence of electron correlations on the
MR at low temperatures.

To confirm this theoretically one could use bosonization
techniques\cite{pereira,hikihara}. However, we will follow another
route based on the ``poor man's'' renormalization method
\cite{matveev1,tsai,devillard}. In our case, the DW scatters  both
the charge and spin of the carriers. As shown below, our scheme
allows us to obtain results for the renormalized transmission and
reflection coefficients in terms of the uncorrelated
spin-dependent ones (i.e., in terms of the reflection and
transmission coefficients of the wall in the absence of
electron-electron interactions). The uncorrelated quantities can
be obtained from other schemes, such as the Hartree-Fock or
density-functional theory (within local density approximation) and
then used as an input in our results to obtain renormalized
transmission through the DW. Hence, our approach -- in combination
with numerical (effective single particle) methods -- offers a new
possibility to understand the material-dependent MR associated
with a DW creation (destruction), and possibly to resolve some
controversy concerning huge magnetoresistance in some ballistic
nanocontacts.

The paper is organized as follows. In Section \ref{scat} we
introduce the problem and the non-interacting scattering  states
for a sharp domain wall. In Section \ref{perturbative} we use
perturbation theory in the electron-electron interaction to
calculate corrections to the scattering amplitudes. We obtain the
renormalization group differential equations for the scattering
amplitudes. In Section \ref{solve} we describe the zero
temperature fixed points predicted by the scaling equations and
the power law behavior of the reflection and transmission
coefficients of the DW as $T\rightarrow 0$. In  Section
\ref{final}  we discuss  the relevance of our findings to
realistic physical systems and summarize our results.

\section{Model}
\label{scat}

We consider a magnetized system with electrons being constrained
to move in one dimension while being  exchange-coupled locally  to
the space  varying magnetization, ${\bf M(r)}$. The wire itself
defines the easy ($z$) axis, and a domain wall centered at $z=0$
separates two regions with opposite magnetizations,
$M_z(z\rightarrow \pm \infty) = \pm M_0$. Assuming  ${\bf M(r)}$
to lie in the $xz$ plane, and the domain wall to be thinner than
the Fermi wavelength, we write the single-particle Hamiltonian as:
\begin{equation}
\hat H_0 = -\frac{\hbar^2}{2m} \frac{d^2}{dz^2} + \hbar V \delta(z)
+J M_z(z)\hat \sigma_z + \hbar \lambda   \delta(z)\hat \sigma_x\,,
\label{model}
\end{equation}
where the term $\hbar \lambda   \delta(z)\hat \sigma_x$ describes
spin scattering produced by the $M_x(z)$ component,\cite{dugaev1}
$$
\lambda = \frac{J}{\hbar}\int_{-\infty}^{\infty} M_x(z)dz\,,
$$
 and $V$ is a potential scattering term.
Single electron wavefunctions are spinors with components
$\chi_\sigma(z)$ satisfying the condition
\begin{equation}
-\frac{\hbar}{2m} \Big(\frac{\partial}{\partial z}\chi_\sigma(0^+) -
\frac{\partial}{\partial z}\chi_\sigma(0^-)\Big)
+V \chi_\sigma(0) - \lambda \chi_{-\sigma}(0) = 0\,.
\label{dois}
\end{equation}
The electron's wavevector in each domain  is related to the energy
$E$ by
\begin{equation}
k=\sqrt{\frac{2m}{\hbar^2}(E\pm JM_0)}\,.
\label{kapa}
\end{equation}
The electron gas in the negative semi-axis ($z<0$) is
predominantly $\up$-spin. An electron incident from the left with
the momentum $k$ and spin $\up$ (or $\dn$) can be transmitted to
the positive semi-axis while preserving its spin, but the energy
conservation requires the momentum to change  from $k$ to $k_-$
(or $k_+$) defined by:
\begin{equation}
k_\pm=\sqrt{k^2 \pm \frac{4m}{\hbar^2}JM_0}\,.
\label{kmn}
\end{equation}
If the transmission occurs with spin reversal, the momentum $k$ is
not changed.

\begin{figure}[ht]
\begin{center}
\epsfxsize=9cm
\epsfbox{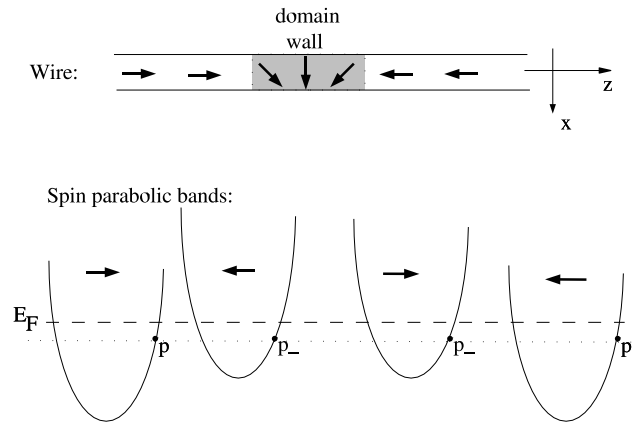}
\end{center}
\caption{Schematic of a domain wall and the relevant electron spin
bands. States $\psi_{p\up}$ and $\psi_{p_-\dn}$ have the same
energy.} \label{fig1}
\end{figure}

We label the states through the incident wave, so that
\begin{equation}
\psi_{k,\up}(z) =  \left(
\begin{array}{c}  e^{ikz}+ r_\up(k)\,  e^{-ikz}\\
r_\up'(k)\,  e^{-ik_{_-}z} \end{array}\right)\,,\qquad z<0
\label{inc}
\end{equation}
describes  a scattering state with a wave incident from
$z=-\infty$ with spin $\up$ and momentum $k>0$. Reflection
amplitudes of a spin $\sigma$ electron with or without spin
reversal are denoted by $r'_\sigma$ and $r_\sigma$, respectively.
The  same convention applies to the transmission amplitudes $
t'_\sigma,\ t_\sigma$. The transmitted wave corresponding to
Eq.~(\ref{inc}) is
\begin{equation}
\psi_{k,\up}(z) =  \left(
\begin{array}{c} t_\up(k)\, e^{ik_{_-}z}\\
t_\up'(k)\,  e^{ikz} \end{array}\right) \,, \qquad z>0
\label{trans}
\end{equation}
and the scattering amplitudes are given by:
\begin{eqnarray}
t_\up(k) &=& \frac{2(v+v_- + 2iV)v }
{ (v+v_- + 2iV)^2 + 4\lambda^2} = r_\up(k) + 1\,,\label{tup}\\
t_\up'(k) &=& \frac{4i\lambda v}{(v+v_- + 2iV)^2
+ 4\lambda^2} = r_\up'(k)\,,\label{tupl}
\end{eqnarray}
where we have defined the velocities $v_{(\pm)} = \hbar k_{(\pm)}/m $.

The scattering state corresponding to a wave incident from the
left with $\dn$-spin is:
\begin{eqnarray}
\psi_{k,\dn}(z<0) &=&  \left( \begin{array}{c} r_\dn'(k) e^{-ik_{_+}z}\\
 e^{ikz}+ r_\dn(k) e^{-ikz} \end{array}\right)\,,
\nonumber \\
\psi_{k,\dn}(z>0) &=&  \left(\begin{array}{c} t_\dn'(k) e^{ikz}\\
t_\dn(k)  e^{ik_{_+}z} \end{array}\right)\,,
\end{eqnarray}
and the corresponding amplitudes are:
\begin{eqnarray}
t_\dn(k) &=& \frac{2(v+v_+ + 2iV)v }{ (v+v_+ + 2iV)^2 + 4\lambda^2} = r_\dn(k)+ 1\,, \label{tdn}\\
t_\dn'(k) &=& \frac{4i\lambda v}{(v+v_+ + 2iV)^2 + 4\lambda^2} = r_\dn'(k)\,.\label{tdnl}
\end{eqnarray}

The expressions for the scattering states corresponding to the
waves incident from $+\infty$ are:
\begin{eqnarray}
\psi_{-k,\up}(z<0) &=& \left(\begin{array}{c}
t_\dn(k) e^{-ik_{_+}z}\\ t_\dn'(k)  e^{-ikz} \end{array}\right)
\nonumber \\
\psi_{-k,\up}(z>0) &=&  \left(\begin{array}{c}
e^{-ikz}+ r_\dn(k)  e^{ikz} \\ r_\dn'(k)  e^{ik_{_+}z} \end{array}\right)
\end{eqnarray}
and
\begin{eqnarray}
\psi_{-k,\dn}(z<0) &=& \left(\begin{array}{c}
t_\up'(k) e^{-ikz}\\ t_\up(k)  e^{-ik_{_-}z} \end{array}\right)
\nonumber \\
\psi_{-k,\dn}(z>0) &=&  \left(\begin{array}{c}
r_\up'(k) e^{ik_{_-}z}\\ e^{-ikz} +  r_\up(k)  e^{ikz}  \end{array}\right)\,,
\end{eqnarray}
where we consider $k>0$. We shall henceforth denote by
$\epsilon(\pm p, \up)$ the eigenenergy of a scattering state with
momentum $+p$ (or $-p$) incident from the left (or right). The
scattering amplitudes satisfy some general relations that can be
found from a generalization of the Wronskian theorem\cite{messiah}
to spinor wavefunctions. We provide such relations in  Appendix
\ref{Wr}.


In order to deal with the electron interactions, it is convenient
to rewrite the scattering states in second quantized form, making
use of right ($\hat a_{q\sigma}$) and left ($\hat b_{q\sigma}$)
moving plane-wave states.

The operators for the scattering states with electrons incident
from the left ($\hat c_{k, \sigma}$) are:
$$
\hat c_{k, \sigma} =\int_{-\infty}^\infty\frac{dq}{2\pi} \left[  \left(
\frac{-i}{q-k-i0}
+ \frac{i t_\sigma^*(k)}{q-k_{-\sigma}+i0} \right)\ \hat a_{q\sigma}\right.
$$
\begin{equation}
- \frac{i r_\sigma^*(k)}{q+k-i0}\ \hat b_{q\sigma}
\left.
+ \frac{i t_\sigma'^*(k)}{q-k+i0}\ \hat a_{q,-\sigma}
- \frac{i r_\sigma'^*(k)}{q+k_{-\sigma}-i0}\ \hat b_{q, -\sigma}\right] \,,
\label{c}
\end{equation}
and  the operators for scattering states with  electrons incident from the
right ($\hat d_{k, \sigma}$) are:
$$
\hat d_{k, \sigma} = \int_{-\infty}^\infty\frac{dq}{2\pi} \left[  \left(
\frac{i}{q+k+i0} - \frac{i t_{-\sigma}^*(k)}{q+k_\sigma-i0} \right)\ \hat b_{q\sigma}
\right.
$$
\begin{equation}
+ \frac{i r_{-\sigma}^*(k)}{q-k+i0}\ \hat a_{q\sigma}
\left.
- \frac{i t_{-\sigma}'^*(k)}{q+k-i0}\ \hat b_{q,-\sigma}
+ \frac{i r_{-\sigma}'^*(k)}{q-k_\sigma+i0}\ \hat a_{q, -\sigma}\right] \,,
\end{equation}
where  $0$ denotes a positive infinitesimal and the $k$-subscript
$\sigma=\pm 1$. By inverting these equations, we obtain the plane
wave operators as linear combinations of the scattering-state operators:
$$
\hat a_{p, \sigma} =\int_{-\infty}^\infty\frac{dk}{2\pi i} \left[
\frac{\hat c_{k, \sigma}}{k-p-i0} -
\frac{ t_{-\sigma}(k) \hat c_{k_+, \sigma}}{k-p+i0}
\right.
$$
\begin{equation}
- \frac{t_{-\sigma}'(k)\hat c_{k, -\sigma}}{k-p+i0}
\left.
- \frac{r_{-\sigma}(k)\hat d_{k,\sigma} }{k-p+i0}\
- \frac{r_{-\sigma}'(k)\hat d_{k_+,-\sigma}}{k-p+i0}\ \right] \,,
\label{a}
\end{equation}
$$
\hat b_{-p, \sigma} =\int_{-\infty}^\infty\frac{dk}{2\pi i} \left[
\frac{\hat d_{k, \sigma}}{k-p-i0} - \frac{ t_\sigma(k) \hat d_{k_-, \sigma}}{k-p+i0}
\right.
$$
\begin{equation}
- \frac{t_\sigma'(k)\hat d_{k, -\sigma}}{k-p+i0}
\left.
- \frac{r_\sigma(k)\hat c_{k,\sigma} }{k-p+i0}\
- \frac{r_\sigma'(k)\hat c_{k_-,-\sigma}}{k-p+i0}\ \right] \,.
\label{b}
\end{equation}

The electron interactions are introduced in the Hamiltonian through the term:
\begin{eqnarray}
\hat H_{int} &=& g_{1,\alpha,\beta}\int
\frac{dk_1dq}{(2\pi)^2}\hat a^\dagger_{k_1,\alpha} \hat b^\dagger_{k_2,\beta}
 \hat a_{k_2+q,\beta} \hat b_{k_1-q,\alpha}\nonumber\\
&+&  g_{2,\alpha,\beta}\int
\frac{dk_1dq}{(2\pi)^2}\hat a^\dagger_{k_1,\alpha} \hat b^\dagger_{k_2,\beta}
 \hat b_{k_2+q,\beta} \hat a_{k_1-q,\alpha} \,.
\label{hint}
\end{eqnarray}
The coupling constants $g_1$ and $g_2$ describe back and forward
scattering processes between opposite moving electrons,
respectively. The Greek letters denote here the spin indices, and
the summation over repeated indices is implied.
 Because the Fermi
momentum depends on spin,  we allow for the dependence of $g$ on
the spins of the interacting particles. We therefore distinguish
between $g_{1\up}$, $g_{1\dn}$  $g_{1\perp}$ and  $g_{2\up}$,
$g_{2\dn}$, $g_{2\perp}$. The forward scattering process between
particles which move in the same direction will not affect the
transmission amplitudes, although it will renormalize the Fermi
velocity\cite{matveev1}. This effect is equivalent to an
effective mass renormalization and the electrons with different spin
orientations may turn out to have different effective masses, in
which case our  calculations remain valid, as shown in  Appendix
\ref{masses}.

\section{Scaling equations}

\label{perturbative} The corrections to the transmission
amplitudes  will be calculated to first order in the perturbation
$\hat H_{int}$. It has been shown in  Ref.~[\onlinecite{matveev1}]
that the corrections  diverge logarithmically  near the Fermi
level. These divergences will later be dealt with in a poor man's
renormalization procedure.

Let us consider the Matsubara propagator,
\begin{equation}
{\cal G}(\tau) =
 -\la T_\tau e^{-\int
 \hat H_{int}(\tau')d\tau'} \hat a_{p, \up} (\tau) \hat c_{p', \up}^\dagger\ra_0\,,
\label{G}
\end{equation}
where $\la ... \ra_0$ denotes the average in the non-interacting Fermi sea.
 The propagator for non-interacting electrons is then given by:
\begin{equation}
{\cal G}^{(0)}(i\omega)  =
\frac{1}{i\omega - \epsilon(p', \up)}\left[
\frac{i}{p-p'+i0} - \frac{i t_\up(p')}{p-p_-'-i0} \right]\,.
\label{zero}
\end{equation}
The transmission amplitude appears associated with the
denominator $p-p_-'-i0$ which, for the variable $p$, gives a pole
in the upper half plane. The  meaning of this pole is that the
transmitted particle is a right-mover in the $z>0$ half-axis. Our
strategy is to calculate the first order correction term (in $\hat
H_{int}$) to ${\cal G}$, which
 will have the same form as the second term in (\ref{zero}), so that
the  amplitude correction,  $\delta t_\up(p')$, can be
 read off from the result. We now explain the procedure in some detail.

We begin by considering the first order expansion for ${\cal G}$
in the coupling $g_{1\up}$. For simplicity, we shall henceforth
omit the  subscript ``0'' in the brackets, since we will be
dealing with  the non-interacting Fermi sea, unless otherwise stated.
From Wick's theorem we get the first order correction to the
propagator  in equation (\ref{zero}) as
$$
{\cal G}^{(1)}(\tau) = g_{1\up} \left[ \right.
$$
$$
 \la \hat a_{p, \up}(\tau) \hat b_{k_2, \up}^\dagger(\tau')\ra
  \la \hat a_{k_1, \up}^\dagger(\tau') \hat b_{k_1-q, \up}(\tau')\ra
 \la \hat a_{k_2+q, \up} (\tau')\hat c_{p', \up}^\dagger\ra
$$
\begin{equation}
\left. +
 \la \hat a_{p, \up}(\tau) \hat a_{k_1, \up}^\dagger(\tau')\ra
  \la \hat b_{k_2, \up}^\dagger(\tau') \hat a_{k_2+q, \up}(\tau')\ra
 \la \hat b_{k_1-q, \up} (\tau')\hat c_{p', \up}^\dagger\ra
\right]
\label{exp1}
\end{equation}
where the internal momenta $k_{1(2)}$, $q$ and time $\tau'$ are to
be  integrated over and the time ordering $T_\tau$ is implicit.
There are also two other Wick paired terms at instant $\tau'$ of
the form $\la \hat a^\dagger(\tau') \hat a(\tau')\ra$   and $\la
\hat b^\dagger(\tau') \hat b(\tau')\ra$. We have omitted these
terms in  Eq.(\ref{exp1}) because they will not be logarithmically
divergent: the divergences arise from electron reflection  by the
Friedel oscillations in the Fermi sea\cite{matveev1}. Such
reflection processes appear in equation (\ref{exp1}) through $\la
\hat b^\dagger(\tau') \hat a(\tau')\ra$ and $\la \hat
a^\dagger(\tau') \hat b(\tau')\ra$.

To calculate $\la \hat a_{k_1, \up}^\dagger\hat b_{k_1-q, \up}\ra$
we make use of the expression (\ref{b}) for $\hat b_{k_1-q, \up}$.
The contour integration over $k_1$ eliminates the terms
containing poles in the same   half-plane. Fermi sea averages,
such as $\la \hat a_{k_1, \up}^\dagger\hat c_{k, \up}\ra$ and $\la
\hat a_{k_1, \up}^\dagger\hat d_{k, \up}\ra$, can be calculated in
the same way as in Eq.~(\ref{zero}). The result is:
$$
\int_{-\infty}^\infty\frac{dk_1}{2\pi} \la
\hat a_{k_1, \up}^\dagger\hat b_{k_1-q, \up}\ra =
$$
\begin{equation}
\int_{-\infty}^\infty\frac{dQ}{2\pi i} \left(
\frac{f(-Q\up) r_\dn^*(Q)}{2Q-q-i0} - \frac{f(Q\up) r_\up^*(Q)}{2Q-q+i0}
\right)\,,
\label{ab}
\end{equation}
where $f(\pm Q\up)$ denotes the Fermi  occupation number of the
state $\psi_{\pm Q,\up}$. In order to calculate the propagator
$-\la \hat T_\tau a_{p, \up}(\tau) \hat b_{k_2, \up}^\dagger\ra$
we again expand $\hat b_{k_2, \up}^\dagger$ using Eq.~(\ref{b}),
and then with help of Eq.~(\ref{rel2}) we obtain
$$
-\int_0^{1/T}d\tau e^{i\omega \tau}\la
\hat T_\tau a_{p, \up}(\tau) \hat b_{-p', \up}^\dagger\ra
=
$$
$$
 \int_{-\infty}^\infty\frac{dQ}{2\pi} \left[ \frac{1}{i\omega-\epsilon(-Q\up)}
\ \frac{1}{Q-p'+i0}\ \frac{r_\dn(Q)}{p-Q-i0}\right.
$$
\begin{equation} \left.
+\   \frac{1}{i\omega-\epsilon(Q\up)}
\ \frac{1}{Q-p'-i0}\ \frac{r_\up^*(Q)}{p-Q+i0} \right]\,.
\label{abdag}
\end{equation}
The presence of two different energy poles  can be understood
from the fact that $\hat a_{q\sigma}$ (or $\hat b_{q\sigma}$)
represents a plane wave running over the entire $z$ axis and its
energy cannot be the same on both sides of the domain wall
because of the energy dependence on spin.

Using  Eqs.~(\ref{ab}) and (\ref{abdag})  we can calculate the
first term in  Eq.~(\ref{exp1}) as
$$
 \la \hat a_{p, \up}(\tau) \hat b_{k_2, \up}^\dagger(\tau')\ra
  \la \hat a_{k_1, \up}^\dagger(\tau') \hat b_{k_1-q, \up}(\tau')\ra
 \la \hat a_{k_2+q, \up} (\tau')\hat c_{p', \up}^\dagger\ra
$$
\begin{eqnarray}
= \frac{1}{i\omega-\epsilon(p'\up)}\int \frac{dQ_1 dQ_2}{(2\pi i)^2}
\frac{1}{i\omega-\epsilon(-Q_2\up)} \nonumber\\
\times\frac{r_\dn(Q_2)r_\dn^*(Q_1)t_\up^*(p') f(-Q_1\up)}{(p-Q_2-i0)(2Q_1-p_-'-Q_2-i0)}\,.
\label{g1up}
\end{eqnarray}
 The analytic continuation of the
Green's function frequency,  $i\omega \rightarrow \omega +i0$,
gives  the retarded Green's function. The frequency denominator
$(i\omega-\epsilon(-Q_2\up))^{-1}$ yields a principal Cauchy part
plus a delta function part. The latter isolates the energy pole at
$\epsilon(-Q_2\up)=\omega$ and we choose
$\omega=\epsilon(p'\up)\Rightarrow Q_2=p_-'$. We shall only retain
this delta function part. Therefore, we put $Q_2=p_-'$ in the
integrand and, by comparing with  (\ref{zero}), we conclude that
the contribution of the first perturbative term in Eq.~(\ref{exp1})
to the transmission amplitude is given by
\begin{equation}
\delta t_\up(p')= -\frac{g_{1\up}}{hv_{F-}}
\int \frac{dQ}{2\pi}\frac{f(-Q\up)}{2Q-2p_-'}
 r_\dn(p_-')r_\dn^*(Q)t_\up^*(p'),
\end{equation}
where $v_{F-}$ now denotes the Fermi velocity corresponding to the
minority  spin Fermi momentum  $ k_{F-}$. A logarithmic divergence
appears as $p'\rightarrow k_{F-}$.

\begin{figure}[ht]
\begin{center}
\epsfxsize=8.5cm
\epsfbox{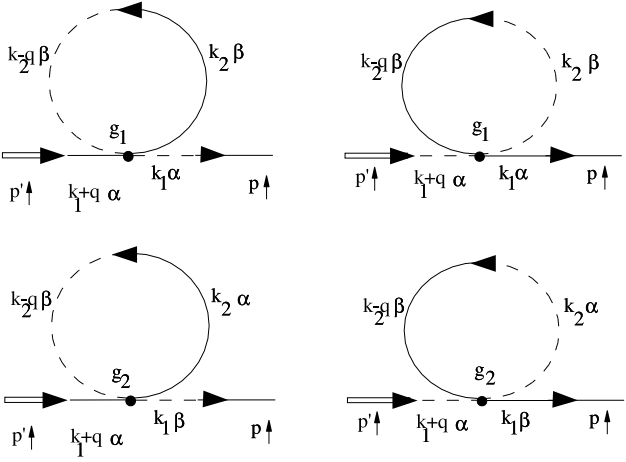}
\end{center}
\caption{Feynmam diagrams for the first order contribution ${\cal
G}^{(1)}$ to the propagator (\ref{G}). The scattering state is
represented by a double line, the $\hat a$ ($\hat b$) particle is
represented by a continuous (dashed) line. The loop represents the
Hartree-Fock potential of the Fermi sea. The scattered electron
exchanges momentum $q$ with the Fermi sea.} \label{fig2}
\end{figure}

The above discussion describes the calculation method. We now need
to calculate all the first order terms in the interactions
$g_{1\alpha\beta}$ and $g_{2\alpha\beta}$. The diagrammatic
representation of   ${\cal G}^{(1)}$ is shown in Fig.~\ref{fig2}.
The horizontal lines represent the electron being scattered  by
the Hartree-Fock potential of the Fermi sea (of scattering
states). Consider, for instance, the upper left diagram: an
electron, initially in state $ c_{p', \up}$ close to the Fermi
level, passes through the barrier as a right-mover ($\hat a$
particle) and then interacts with the Fermi sea (on the positive
$z$ semi-axis). The electron is reflected (from $\hat a$ to $\hat
b$ particle) while  exchanging momentum $q$ with the Fermi sea.
Finally, it is reflected by the barrier again, becoming a spin-up
right mover with momentum  $p$. A logarithmic divergence occurs if
the polarized Fermi sea can provide exactly the momentum that is
required to keep the  electron  always near the Fermi level during
the intermediate virtual steps. Concerning the spin dependence of
the interaction parameters, we distinguish between $g_{1\up}$,
$g_{2\up}$, which describe interaction between spin majority
particles (that is spin-$\up$ on the right and spin-$\dn$ on the
left of the barrier) and $g_{1\dn}$, $g_{2\dn}$, which describe
interaction between spin minority particles (that is spin-$\dn$ on
the right and  spin-$\up$ on the left of the barrier). We use
$g_{1\perp}$, $g_{2\perp}$ to denote interaction between particles
with opposite spins. According to the physical interpretation of
the Feynman diagrams just given above, we always know on which side of
the barrier the interaction with the Fermi sea (closed loop in the
diagram) is taking place.

It can be seen that the $g_{1\perp}$ terms are proportional to
$\log|k_{F+}-k_{F-}|$, so they do not diverge. The logarithmic
divergence would be restored in a spin degenerate system
($k_{F+}=k_{F-}$). This can be understood from the diagrams in
Fig.~\ref{fig2} as follows: the electron with spin $\alpha$ is
reflected by a polarized Fermi sea with spin $-\alpha$. The
momentum provided by the Fermi sea is $2k_{F-\alpha}$, while the
momentum required by the electron is  $2k_{F\alpha}$. The
$g_{2\perp}$ terms  produce logarithmic divergences that would not
exist in the absence of spin-flip scattering ($t'=r'=0$).
Introducing the Fermi level velocities $v_{F\pm}$ for majority or
minority spin particles, we write the diverging contributions to
$\delta t_\up(p')$  as
\begin{eqnarray}
\delta t_\up(p')&=&
\int^{K_{F-}}\frac{dQ}{4hv_{F-}} \frac{
(g_{2\dn}-g_{1\dn})r_\dn(p_-')r_\dn^*(Q)t_\up(p')}{Q-p_-'} \nonumber\\
&+& \int^{K_{F+}}\frac{dQ}{4hv_{F+}}
\frac{(g_{2\up}-g_{1\up})t_\up(p')r_\up^*(Q)r_\up(p')}{Q-p'} \nonumber\\
&+&  \int^{K_{F+}}\frac{dQ}{4hv_{F+}}
\frac{(g_{2\up}-g_{1\up})r_\up'(p')r_\up^*(Q)t_\up'(p')}{Q-p'}  \nonumber\\
&+& \int^{K_{F-}}\frac{dQ}{4hv_{F-}} \frac{(g_{2\dn}-g_{1\dn})
t_\dn'(p'-)r_\dn^*(Q)r_\up'(p')}{Q-p_-'}  \nonumber\\
&+& \int^{K_{F+}}\frac{dQ}{2hv_{F+}} \frac{g_{2\perp}
r_\up'^*(Q)t_\up(p')r_\up'(p')}{Q+Q_{-}-p'-p_-'} \nonumber\\
&+& \int^{K_{F+}}\frac{dQ}{2hv_{F-}} \frac{g_{2\perp}
r_\up'^*(Q)t_\dn'(p'_-)r_\up(p')}{Q+Q_{-}-p'-p'_-} \nonumber\\
&+&\int^{K_{F+}}\frac{dQ}{2hv_{F+}} \frac{g_{2\perp}
r_\up'^*(Q)r_\up'(p')t_\up(p')}{Q+Q_{-}-p'-p_-'} \nonumber\\
&+& \int^{K_{F-}}\frac{dQ}{2hv_{F-}} \frac{g_{2\perp}
r_\dn'^*(Q)r_\dn(p_-')t_\up'(p')}{Q+Q_{+}-p'-p_-'}\,,
\label{ints}
\end{eqnarray}
where $Q_{\pm}$ is  related to $Q$ as in Eq.~(\ref{kmn}). In order
to apply the poor man's renormalization method, it is preferable
to transform the momentum integrations in Eq.~(\ref{ints}) into
energy integrals. In order to do this, we linearize the spectrum
near the Fermi level as:
\begin{eqnarray}
\hbar v_{F+} (Q-K_{F+}) &=& \hbar v_{F-} (Q_--K_{F-}) \equiv
\epsilon ,\\ \hbar v_{F+} (p'-K_{F+}) &=& \hbar v_{F-}
(p_-'-K_{F-}) \equiv \epsilon' ,
\end{eqnarray}
where energy of the scattered electron is $\epsilon'$ and the
energies $\epsilon (\epsilon^{'})<0$ are measured with respect to
the Fermi level. The linearization is assumed to be valid within
an energy range $D$ around the Fermi level. The $Q-$integrals
appearing in equation (\ref{ints}) can now be written as:
\begin{eqnarray}
\int^{K_{F-}}\frac{dQ}{Q-p_-'} &=& \int_{-D}^0
\frac{d\epsilon}{\epsilon-\epsilon'}\; ,
\nonumber\\
\int^{K_{F+}}\frac{dQ}{Q-p'} &
=& \int_{-D}^0 \frac{d\epsilon}{\epsilon-\epsilon'}\; ,
\nonumber\\
\int^{K_{F\mp}}\frac{dQ}{Q+Q_{\pm}-p'-p_-'} &=&
\frac{v_{F\pm}}{v_{F+} + v_{F-}}\int_{-D}^0
\frac{d\epsilon}{\epsilon-\epsilon'}\; .
\nonumber
\end{eqnarray}
The scattering amplitudes with $\up$ ($\dn$) spin index are always
associated with the momentum $p'$ ($p_-'$). Therefore, we shall
henceforth omit the momentum argument $p'$ ($p_-'$) of the
scattering amplitudes. The divergent perturbative correction,
$\delta t_\up$, is  proportional to $\log (|\epsilon'|/D)$,
\begin{eqnarray}
\frac{\delta t_\up}{\log\frac{|\epsilon'|}{D}} &=&
 \frac{(g_{2\dn}-g_{1\dn})}{4hv_{F-}}   r_\dn^* r_\dn t_\up
\ +\
\frac{(g_{2\up}-g_{1\up})}{4hv_{F+}} r_\up^* r_\up  t_\up   \nonumber\\
&+&
\frac{(g_{2\up}-g_{1\up})}{4hv_{F+}}   r_\up^*  r_\up' t_\up' \ +\
\frac{(g_{2\dn}-g_{1\dn})}{4hv_{F-}} r_\dn^* r_\up' t_\dn' \nonumber
\end{eqnarray}
$$
+\frac{g_{2\perp}}{2h(v_{F+} + v_{F-})} \left(
r_\dn'^*  r_\up' t_\up\  +\
r_\up'^*  r_\dn t_\up' \right.
$$
\begin{equation}
\left. +\ r_\up'^* r_\up t_\dn'\ + \
r_\dn'^* r_\up' t_\up\  \right)  \,.
\end{equation}

For the  calculation of $\delta t_\up'(p')$ and  $\delta
t_\dn(p'_-)$, the propagators we need to consider are $ -\la
T_\tau \hat a_{p, \dn}(\tau)  \hat c_{p', \up}^\dagger\ra$ and $
-\la T_\tau \hat a_{p, \dn}(\tau)  \hat c_{p_-', \dn}^\dagger\ra$,
respectively.
 The perturbation theory  is analogous to that described above.
In order to obtain $t_\dn'(p'_-)$ we consider the propagator
$ -\la T_\tau \hat a_{p, \up}(\tau)  \hat c_{p_-', \dn}^\dagger\ra$.

The logarithmically divergent perturbative terms can be dealt with
using a renormalization procedure: we reduce  the bandwidth $D$ by
removing  states in a narrow strip $\delta D$ near the band edge
and work again the perturbation theory in the new bandwidth
$D-\delta D$. The effect of removal of the band edge  states must
be compensated  by adopting  a new value of  $t_\up$ for the new
bandwidth. Applying this  procedure  step by step yields
successive renormalizations of  $t_\up$. A differential equation
is obtained by noting that the perturbation theory result, $t_\up+
\delta t_\up$,  remains invariant as $D$ is reduced:
$$
dt_\up \ + \ \frac{\partial \delta t_\up}{\partial D}\ dD = 0\,.
$$
We introduce now a variable $\xi = \log (D/D_0)$ which will be
integrated from $0$ to $\log (|\epsilon'|/D_0)$, corresponding to
the fact that the bandwidth is progressively reduced from $D_0$ to
$|\epsilon'|$ (which will eventually be taken as temperature:
$|\epsilon'|=T$) and the scaling differential equations for the
transmission amplitudes become:
\begin{eqnarray}
\frac{d t_\up}{d\xi} &=&
 \frac{(g_{2\dn}-g_{1\dn})}{4hv_{F-}} \left[ \
 r_\dn^* r_\dn t_\up + r_\dn^* r_\up' t_\dn'\ \right]  \nonumber\\
&+&
\frac{(g_{2\up}-g_{1\up})}{4hv_{F+}}  \left[ \
r_\up^* r_\up  t_\up + r_\up^*  r_\up' t_\up'\  \right] \nonumber\\
&+&
\frac{g_{2\perp}}{2h(v_{F+} + v_{F-})} \left[
r_\dn'^*  r_\up' t_\up\  +\
r_\up'^*  r_\dn t_\up'\ \right. \nonumber\\
 &+& \left. r_\up'^* r_\up t_\dn' + r_\dn'^* r_\up' t_\up\  \right]\,,
\label{tupdif}
\end{eqnarray}
\begin{eqnarray}
\frac{d t_\up'}{d\xi} &=&
 \frac{(g_{2\dn}-g_{1\dn})}{2hv_{F-}}   r_\dn^*  r_\dn' t_\up \nonumber\\
 &+& \frac{(g_{2\up}-g_{1\up})}{2hv_{F+}}   r_\up^* r_\up t_\up'\nonumber\\
&+&
\frac{g_{2\perp}}{h(v_{F+} + v_{F-})} \left[
r_\dn'^* r_\up t_\up\ + \
r_\up'^* r_\dn' t_\up'\ \right]\,,
\label{tpupdif}
\end{eqnarray}
\begin{eqnarray}
\frac{d t_\dn}{d\xi} &=&
 \frac{(g_{2\up}-g_{1\up})}{4hv_{F+}} \left[ \
 r_\up^* r_\up t_\dn\  +\ r_\up^* r_\dn' t_\up'\ \right]  \nonumber\\
&+&
\frac{(g_{2\dn}-g_{1\dn})}{4hv_{F-}}  \left[ \
r_\dn^* r_\dn  t_\dn +   r_\dn^*  r_\dn' t_\dn' \ \right] \nonumber\\
&+&
\frac{g_{2\perp}}{2h(v_{F+} + v_{F-})} \left[
r_\up'^*  r_\dn' t_\dn\  +\
r_\dn'^*  r_\up t_\dn'\ \right. \nonumber\\ &+& \left.
r_\dn'^* r_\dn t_\up'\ + \
r_\up'^* r_\dn' t_\dn\  \right]\,,
\label{tdndif}
\end{eqnarray}
\begin{eqnarray}
\frac{d t_\dn'}{d\xi} &=&
 \frac{(g_{2\up}-g_{1\up})}{2hv_{F+}}   r_\up^*  r_\up' t_\dn\ + \
\frac{(g_{2\dn}-g_{1\dn})}{2hv_{F-}}   r_\dn^* r_\dn t_\dn'\nonumber\\
&+&
\frac{g_{2\perp}}{h(v_{F+} + v_{F-})} \left[
r_\up'^* r_\dn t_\dn\ + \
r_\dn'^* r_\up' t_\dn'   \right]\,.
\label{tpdndif}
\end{eqnarray}

In order to obtain the perturbative correction to the reflection amplitude
$r_\up (p')$ we consider the propagator:
\begin{eqnarray}
 {\cal G}(\tau)&=& -\la T_\tau \hat b_{p, \up}(\tau)
 \hat c_{p', \up}^\dagger\ra \  \Rightarrow\nonumber\\
{\cal G}^{(0)}(i\omega)  &=&
\frac{1}{i\omega - \epsilon(p', \up)}\  \frac{i r_\up(p')}{p+p'+i0} \,.
\label{rup}
\end{eqnarray}
In this case,  there is a process where the incoming electron from
the  left is reflected back by the Hartree potential without even
crossing the domain wall. The corresponding   term comes from the
Wick pairing term
$$
\ \la \hat b_{p, \up}(\tau) \hat b_{k_2, \up}^\dagger(\tau')\ra
\  \la \hat a_{k_1, \up}^\dagger(\tau') \hat b_{k_1-q, \up}(\tau')\ra
\ \la \hat a_{k_2+q, \up} (\tau')\hat c_{p', \up}^\dagger\ra
$$
and gives a contribution to $\delta  r_\up (p')$ equal to
$$
\frac{g_{2\up}-g_{1\up}}{4hv_{F+}}r_\up(p') \log\frac{|\epsilon'|}{D}\,.
$$
The differential equation for  $r_\up (p')$ is
\begin{eqnarray}
\frac{ d r_\up}{d\xi} &=&
 \frac{g_{2\up}-g_{1\up}}{4hv_{F+}}\left[
 r_\up^*  r_\up  r_\up\  +\  r_\up^* t_\up't_\up' \right] \nonumber\\ &+&
\frac{g_{2\dn}-g_{1\dn}}{4hv_{F-}} \left[
r_\dn^*  t_\up  t_\dn\  +r_\dn^* r_\dn'  r_\up'\  \right]\nonumber\\ &-&
\frac{g_{2\up}-g_{1\up}}{4hv_{F+}}r_\up  \nonumber\\
&+& \frac{g_{2\perp}}{2h(v_{F+} + v_{F-})} \left[ \  r_\up'^*
r_\dn' r_\up\ + r_\dn'^* r_\up' r_\up\  \right. \nonumber\\ &+&
\left. r_\up'^* t_\dn t_\up'\ +\ r_\dn'^* t_\up t_\up'\
\right], \label{ruplogs}
\end{eqnarray}
and the differential equation for  $r_\up' (p')$ is
\begin{eqnarray}
\frac{ d r_\up'}{d\xi} &=&
 \frac{g_{2\up}-g_{1\up}}{4hv_{F+}} \left[
 r_\up^*  r_\up  r_\up'\  +\ r_\up^* t_\up't_\up\ \right]  \nonumber\\ &+&
\frac{g_{2\dn}-g_{1\dn}}{4hv_{F-}} \left[
r_\dn^*  t_\up  t_\dn'\  + r_\dn^* r_\dn  r_\up'\  \right]
  \nonumber\\
&+& \frac{g_{2\perp}}{2h(v_{F+} + v_{F-})} \left[ \  r_\up'^*
r_\dn r_\up\ + r_\dn'^* r_\up' r_\up'\  \right. \nonumber\\  &+&
\left. r_\up'^* t_\dn' t_\up'\ +\ r_\dn'^* t_\up t_\up\
\right] - \frac{g_{2\perp}}{2h(v_{F+} + v_{F-})} r_\up'.
\label{rupup}
\end{eqnarray}

\section{Fixed points}
\label{solve}

The parameters of the model, which enter the scaling equations
are:
\begin{eqnarray}
 \frac{g_{2\up}-g_{1\up}}{4hv_{F+}} &=& g_\up\,, \\
\frac{g_{2\dn}-g_{1\dn}}{4hv_{F-}} &=& g_\dn\,,\\
\frac{g_{2\perp}}{2h(v_{F+} + v_{F-})} &=& g_\perp\,,
\end{eqnarray}
and the ratio $v_{F-}/v_{F+}$. The results can be presented in
terms of the transmission coefficients, defined by
\begin{eqnarray}
 \cal T_\up &=& \frac{v_{F-}}{v_{F+}} |t_\up|^2\,, \\
  \cal T_\up' &=& |t_\up'|^2\,, \\
 \cal T_\dn' &=& |t_\dn'|^2\,,
\end{eqnarray}
and the reflection coefficients
\begin{eqnarray}
 \cal R_\up &=&  |r_\up|^2 \,,\\
 \cal R_\dn &=& |r_\dn|^2\,, \\
\cal R_\up' &=&  \frac{v_{F-}}{v_{F+}} |r_\up'|^2\,. \label{Rpdef}
\end{eqnarray}
By definition, these coefficients refer to the respective
currents divided by the incident current.

\subsection{Insulator fixed points}
\label{3conditions}

\begin{figure}[ht]
\begin{center}
\epsfxsize=8cm
\epsfbox{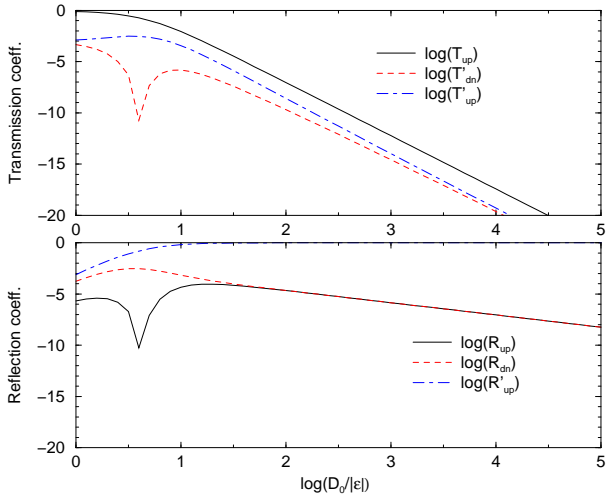}
\end{center}
\caption{Logarithm of transmission/reflection coefficients versus
$\log\frac{D_0}{|\epsilon|}$. We may identify temperature $T$ with
$|\epsilon|$. The interaction parameters are: $g_\up=g_\dn=1$,
$g_\perp=1.3$ and the noninteracting domain wall model parameters
are $V=0$, $v_{F-}/v_{F+} = 0.8$,  $v_{F+}=1$,  $\lambda=0.2$. The
dips are due to the sign reversal of the (small) scattering
amplitudes. The long linear tails are analytically  described in
the text. At low temperature the system becomes a 100\% spin flip
reflector.} \label{strange}
\end{figure}

\begin{figure}[ht]
\begin{center}
\epsfxsize=8cm
\epsfbox{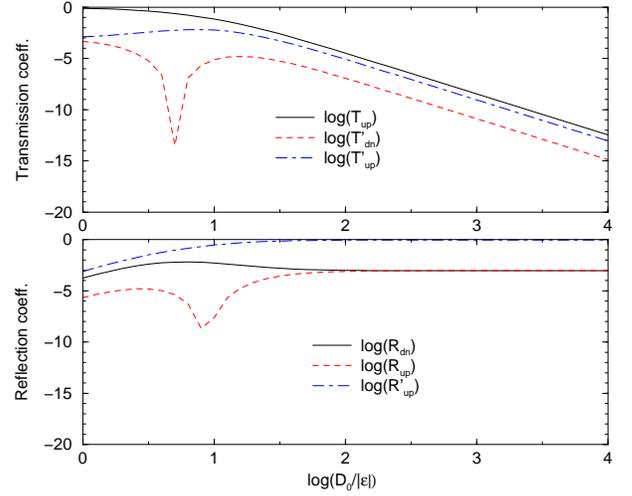}
\end{center}
\caption{Logarithm of the transmission/reflection coefficients
versus $\log (D_0/|\epsilon|)$. We may identify temperature $T$
with $|\epsilon|$. The parameters are:$V=0$, $(v_{F-}/v_{F+}) =
0.8$, $g_\up=g_\dn=g_\perp=1$, $v_{F+}=1$, $\lambda=0.2$. The
reflection coefficients are all finite as $T\rightarrow 0$.}
\label{figscale}
\end{figure}

We have made a numerical study of the scaling equations. The
non-interacting domain wall  described in Section \ref{scat}
provides the  initial scattering parameters for our numerical
scaling. Below we describe analytically the scaling behavior close
to the fixed points we have found.

For repulsive interactions ($g_\up, g_\dn, g_\perp>0$) the system
flows to insulator fixed points. For a moderate to large
$\lambda/v_{F+}$ (larger than about $0.1$) all the transmission
amplitudes, $t_\sigma$ and $t_\sigma'$, vanish faster than any
reflection amplitude as $T\rightarrow 0$. We may then rewrite the
scaling equations neglecting the small transmission amplitudes.
The scaling equation for $r_\up$, for instance, becomes
\begin{equation}
\frac{ d r_\up}{d\xi} =
 g_\up \left(\   |r_\up|^2  -1\ \right) r_\up\   +\
g_\dn\    r_\dn^* r_\dn'  r_\up'\  +
2g_\perp\ \frac{v_{F-}}{v_{F+}} |r_\up'|^2 r_\up \,,
\label{newrup}
\end{equation}
where we used  Eq.~(\ref{a10}). The Wronskian relation (\ref{a6}),
allowing for complex reflection amplitudes, shows that
\begin{equation}
r_\dn^* r_\dn' +  r_\up r_\dn'^* = 0\,.
\label{a6insulator}
\end{equation}
The charge conservation condition is satisfied solely by the
reflections,
\begin{equation}
1\ =\ |r_\up|^2 + \frac{v_{F-}}{v_{F+}} |r_\up'|^2\
=\   |r_\dn|^2 + \frac{v_{F-}}{v_{F+}} |r_\up'|^2\,,
\label{a12ins}
\end{equation}
from which we easily conclude that $ |r_\up| =  |r_\dn|$ at the
fixed point. Then, Eq.~(\ref{newrup}) may be rewritten as
\begin{eqnarray}
\frac{ d r_\up}{d\xi} &=&  \frac{v_{F-}}{v_{F+}} \
\left(\ 2g_\perp - g_\up-g_\dn  \right)  |r_\up'|^2\ r_\up \nonumber\\ &=&
 \frac{v_{F-}}{v_{F+}} \  \left(\ 2g_\perp - g_\up-g_\dn  \right)
 \left(\  1-|r_\up|^2 \right) r_\up .
\label{finalrup}
\end{eqnarray}

Consider now the scaling equation (\ref{rupup}). For $r_\up'$ in case of
negligible transmissions we have
\begin{eqnarray}
\frac{ d r_\up'}{d\xi} &=& g_\up  |r_\up|^2  r_\up'\  +\ g_\dn
|r_\dn|^2  r_\up' \nonumber\\ &+& g_\perp  \left(\  r_\up'^* r_\dn
r_\up\  + \frac{v_{F-}}{v_{F+}} |r_\up'|^2 r_\up' - r_\up'\right)
.\label{ruppins}
\end{eqnarray}
The Wronskian relation (\ref{a6}), allowing for complex reflection amplitudes, tell us that
\begin{equation}
r_\dn r_\up'^* +  \frac{v_{F-}}{v_{F+}}r_\up^* r_\dn' = 0,
\label{a6ins}
\end{equation}
and we recast Eq.~(\ref{ruppins}) as
\begin{equation}
\frac{ d r_\up'}{d\xi} = \left(\  g_\up + g_\dn - 2 g_\perp\right)
 \left( 1-\frac{v_{F-}}{v_{F+}} |r_\up'|^2 \right) r_\up'.\
\label{ruppfinal}
\end{equation}
In the derivation of (\ref{finalrup}) and (\ref{ruppfinal}) the
only  assumption made was that the transmission amplitudes are
negligibly small. The reflection amplitudes may be, in general, complex
and are still renormalized  after the transmissions became
negligible.

Now we see that Eqs.~(\ref{finalrup}) and (\ref{ruppfinal}) predict
that the phases of the complex numbers $r_\up$,  $r_\up'$ are
unchanged during scaling. The two fixed points we may consider
correspond to  $r_\up$ approaching 0, or $|r_\up|$ approaching 1
along a constant phase line  in the complex plane.

The situation  $|r_\up|\rightarrow 0$ requires $2g_\perp -
g_\up-g_\dn >0$ and, by charge conservation we have
$|r_\up'|\rightarrow \sqrt{v_{F+}/v_{F-}}$. Upon integrating
(\ref{ruppfinal}) with $\xi$ ranging from $0$ to $\log(T/D_0)$,
the amplitude $r_\up'$ will vary from its
 initial value $r_{\up,0}'$ to $r_\up'(T)$.
Using the definition   (\ref{Rpdef}) for the reflection
coefficient, we write
\begin{equation}
{\cal R}_\up'(T)=\frac{\frac{{\cal R}_{\up,0}'}{1-{\cal R}_{\up,0}'}
\left(\frac{T}{D_0}\right)^{2\left(g_\up+g_\dn-2g_\perp\right)}}
{ 1+ \frac{{\cal R}_{\up,0}'}{1-{\cal R}_{\up,0}'}
\left(\frac{T}{D_0}\right)^{2\left(g_\up+g_\dn-2g_\perp\right)}}\,.
\label{sfins}
\end{equation}
If  $2g_\perp - g_\up-g_\dn >0$, then ${\cal R}_\up'(T)\rightarrow
1$  as $T\rightarrow 0$. The domain wall becomes insulating. It
reflects all incident electrons while reversing their spin.
Therefore, such a DW may be considered as a perfect spin-flip
reflector at zero temperature. In order to find the low $T$
behavior of transmissions we put $r_\up=r_\dn=0$ in equations
(\ref{tupdif})-(\ref{tpdndif}) and obtain
\begin{equation}
|t_\up| \sim |t_\up'|\sim |t_\dn'| \sim T^{2g_\perp} \,.
\end{equation}
Figure~\ref{strange} shows numerical solutions to the scaling
equations, where the system is flowing to this fixed point.

In the regime where $g_\up+g_\dn-2g_\perp >0$ we have ${\cal
R}_\up'(T)\rightarrow 0$,  ${\cal R}_\up(T)\rightarrow 1$. So,
the domain wall reflects all incident electrons while  preserving
their spin. From Eqs.~(\ref{tupdif})-(\ref{tpdndif}) for the
transmission amplitudes we obtain:
\begin{eqnarray}
|t_\up| &\sim& T^{g_\up + g_\dn},
\nonumber \\
|t_\up'| &\sim&  T^{2g_\up},
\nonumber\\
|t_\dn'| &\sim& T^{2g_\dn} .
\label{tvanish}
\end{eqnarray}

If $g_\up+g_\dn-2g_\perp =0$ then both  $\cal R_\up'(T)$ and $\cal
R_\up(T)$ tend to finite values. Such a regime is illustrated in
Fig.~\ref{figscale}. In this case, Eqs.~(\ref{tupdif}),
(\ref{tpupdif}) and (\ref{tpdndif}) with constant reflection
amplitudes become a linear (in $t_\up$, $t_\up'$, $t_\dn'$)
algebraic $3\times 3$ system. The eigenvalues of the matrix give three
temperature exponents and each transmission amplitude will be a
linear combination of the three powers of $T$. For decreasing
temperature, there may be a crossover from one exponent to the other
and the lowest  exponent dominates as $T\rightarrow 0$.

For smaller values of $\lambda/v_{F+}$ (smaller than about $0.1$)
in the Hamiltonian (\ref{model}), the system flows to a fixed
point, where $r_\up'$ vanishes faster than  the transmissions and
$|r_\sigma|\rightarrow 1$.  The transmission amplitudes still
scale to zero as in Eqs.~(\ref{tvanish}). The  scaling equation
(\ref{rupup}) for $r_\up'$ can be linearized in $r_\up'$ by
neglecting the second order terms in  $t$, $t'$, and considering
that $|r_\sigma |\rightarrow 1$,
\begin{eqnarray}
\frac{d r_\up'}{d\xi} &=& \left( g_\up + g_\dn \right) r_\up',
\end{eqnarray}
from which we obtain
\begin{equation}
|{\cal R}_\up'| \sim T^{2\left(g_\up + g_\dn\right)} .
\end{equation}
We see that the exponent for $|r_\up'(T)|$ is not bigger than the
exponents in (\ref{tvanish}). Now, in the scaling equation
(\ref{ruplogs}) for $r_\up$ we cannot neglect the terms containing
transmission amplitudes on the right hand side. One can easily see
that the $g_\up$ term becomes $g_\up(|r_\up|^2 - 1)r_\up$, which
is of the same order of magnitude as the other terms.
Consequently, the scaling behavior derived in Eqs.~(\ref{finalrup})
and (\ref{ruppfinal}) does not apply here, since it was assumed
there that the transmissions were smaller than $r_\sigma'$. The
behavior of $r_\sigma$ as $T\rightarrow 0$ can be found from the
charge conservation condition: ${\cal R}_\up = 1- {\cal R}_\up' -
{\cal T}_\up - {\cal T}_\up'$, so $1- {\cal R}_\up \sim T^{{\rm
min}\left\{2(g_\up + g_\dn), 4g_\up \right\}}$.
\begin{figure}[ht]
\begin{center}
\epsfxsize=9cm
\epsfbox{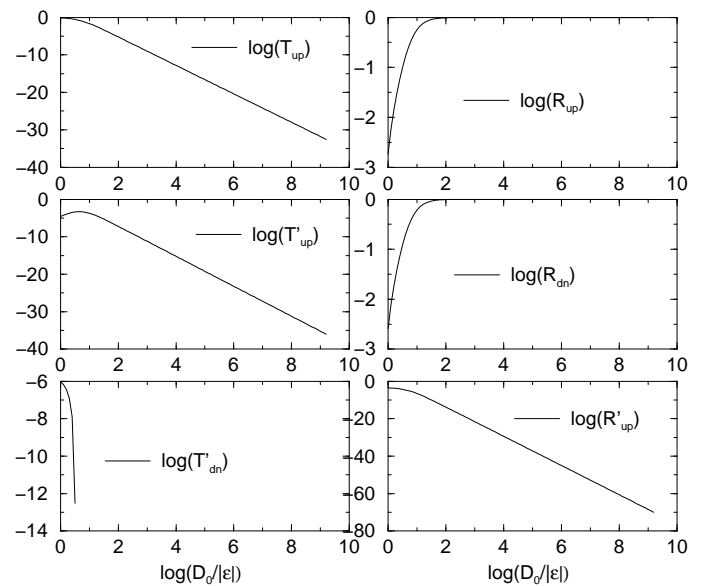}
\end{center}
\caption{Logarithm of transmission/reflection coefficients versus
$\log (D_0/|\epsilon|)$. We may identify temperature $T$ with
$|\epsilon|$. Parameters are: $g_\up=1$, $g_\dn=0.9$,
$g_\perp=1.3$ and $v_{F-}/v_{F+} = 0.8$. For the initial
noninteracting scattering amplitudes we used $V=0$, $v_{F+}=1$,
$\lambda=0.05$.} \label{smalllamb}
\end{figure}
Such a situation is shown in Fig.~\ref{smalllamb}, where $t_\dn'$
is seen to initially flow very fast  to zero. The explanation is
the following: for small $\lambda$ in Eqs.~(\ref{tup}) and
(\ref{tupl}), the noninteracting domain wall has $t_\up>1$,
$r_\up>0$ and  $t_\dn<1$, $r_\dn<0$. Also $t'=r'$ is small. The
scaling equation for  $t_\dn'$  becomes
$$
\frac{d t_\dn'}{d\xi} = \left( 2g_\up r_\up  + 2g_\perp |r_\dn|
\right)  r_\dn' t_\up + \left( 2g_\up r_\dn^2 - 2g_\perp |r_\up'
r_\dn'| \right)t_\dn' .
$$
The first term on the right-hand side is positive and much larger
than the second one, so $t_\dn'$  tends fast to zero
 and disappears from the equations.
The equation for  $t_\up'$ is
$$
\frac{d t_\up'}{d\xi} = \left( 2g_\dn r_\dn - 2g_\perp r_\up
\right) r_\dn t_\up + \left( 2g_\up r_\up^2 + 2g_\perp |r_\up'
r_\dn'| \right) t_\up' .
$$
The first term on the right is negative while the second is
smaller because of small initial $t_\up'$. Then,  $t_\up'$
initially grows as can be seen in Fig.~\ref{smalllamb}.

\subsection{Transparent barrier fixed points}

\begin{figure}[ht]
\begin{center}
\epsfxsize=8.5cm
\epsfbox{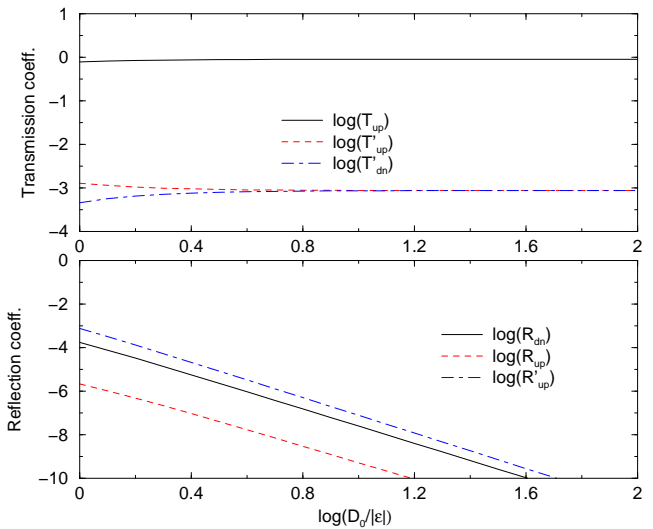}
\end{center}
\caption{Logarithm of transmission/reflection coefficients versus
$\log (D_0/|\epsilon|)$. Parameters are: $g_\up= -0.7$, $g_\dn=-
1.1$, $g_\perp= -1$ and $v_{F-}/v_{F+} = 0.8$. For the initial
noninteracting scattering amplitudes we used $V=0$, $v_{F+}=1$,
$\lambda= 0.2$. The transmission coefficients are all finite as
$T\rightarrow 0$.} \label{atractive1}
\end{figure}

Zero temperature fixed points corresponding to a transparent
domain wall can be achieved when the interaction constants are all
negative, i.e., for attractive electron interaction. Although we
do not expect such a situation to occur in realistic physical
systems, we describe below the fixed points for the case $V=0$ in
the model Hamiltonian (\ref{model}). For moderate to strong
$\lambda/v_{F+}$ in the model (\ref{tup})-(\ref{tupl}), the zero
temperature values $1\ge |t_\up'|=|t_\dn'|>|t_\up|$ depend on the
initial parameters. Smaller  $\lambda/v_{F+}$ enhances $t_\up$
relative to $t_\sigma'$. The reflection coefficients vanish under
scaling as powers of temperature. The corresponding exponents can
be obtained after linearizing (for small reflections) the scaling
equations (\ref{ruplogs})-(\ref{rupup}). The resulting 3 by 3
matrix contains the  finite  limiting values of the transmission
amplitudes and its eigenvalues give the temperature exponents for
the vanishing reflection amplitudes. Figure \ref{atractive1} shows
an example of this behavior.

If some of the interaction constants are positive and the others
negative, the situation becomes more complex. Below we describe
several possible situations.

\subsubsection{The case $g_\up$, $g_\dn>0$,  $g_\perp<0$}

The system flows to the fixed point $r_\up=r_\dn=-1$ with all
other amplitudes vanishing. The low-$T$ behavior of the
transmission can be easily found by inserting the fixed point
reflections into Eqs.~(\ref{tupdif})-({\ref{tpdndif}):
\begin{equation}
|t_\up| \sim T^{g_\up + g_\dn}\,,\qquad
|t_\up'| \sim  T^{2g_\up} \,,\qquad
 |t_\dn'| \sim T^{2g_\dn}\,.
\end{equation}
The scaling equation for $r_\up'$, neglecting second order terms
in the scattering amplitudes, takes the form:
\begin{equation}
\frac{ d r_\up'}{d\xi} = \left(\  g_\up + g_\dn - 2 g_\perp\right)
r_\up'\ \Rightarrow\  r_\up' \sim T^{g_\up+g_\dn-2g_\perp}\,,
\end{equation}
so that we must have $g_\up + g_\dn - 2 g_\perp >0$ in order for $ r_\up'\ \rightarrow 0$.

\subsubsection{The case $g_\up$, $g_\dn<0$,  $g_\perp>0$}

The system flows to the perfect spin-flip reflector fixed point
$|r_\up'|=\sqrt{v_{F+}/v_{F-}}$ with all other amplitudes
vanishing, in accordance with the condition $g_\up + g_\dn - 2
g_\perp <0$ derived earlier.

\subsubsection{The case $g_\up>0$, $g_\dn<0$}

For a negative or small positive $g_\perp$, the system flows to a
fixed point where $|t_\dn'|=1$, $r_\up=-1$. The wall transmits all
spin-$\dn$ particles with a spin-flip and reflects all spin-$\up$
particles. From Eqs.~(\ref{tupdif})-({\ref{tpdndif}) we see that
the exponents for the transmission amplitudes are:
\begin{equation}
|t_\up| \sim T^{g_\up}\,,\qquad
|t_\up'| \sim  T^{2g_\up}\,.
\end{equation}
After linearizing Eq.~(\ref{rupup}) for $r_\up'$ in small
amplitudes, we have
\begin{equation}
\frac{ d r_\up'}{d\xi} \approx
\left( g_\up - g_\perp\right) r_\up' \ \Rightarrow\ r_\up' \sim T^{ g_\up - g_\perp}\,,
\label{rluplin}
\end{equation}
which requires $g_\up > g_\perp$ for vanishing $r_\up'$. If $g_\up
- g_\perp$ is small, the small quantities neglected in the
right-hand-side of Eq.~(\ref{rluplin}) become important. Therefore,
this fixed point holds for $g_\up - g_\perp$ above some small
quantity. Linearizing Eq.~(\ref{rup}) for $r_\up$ in small
amplitudes, we find
\begin{equation}
\frac{ d r_\dn}{d\xi} \approx
- g_\dn r_\dn^*  - g_\dn r_\dn  \ \Rightarrow\  r_\dn \sim T^{-2g_\dn}\,,
\end{equation}
which tends to zero since $g_\dn <0$. For larger  $g_\perp$  the
system flows to the spin-flip reflector fixed point ($ |r_\up'|
\rightarrow \sqrt{v_{F+}/v_{F-}}$).

\subsubsection{The case $g_\up<0$, $g_\dn>0$}

The situation is analogous to the previous one. For negative or
small positive $g_\perp$ the system flows to a fixed point where
$|t_\up'|=1$, $r_\dn=-1$ with all the others vanishing. The wall
transmits all spin-$\up$ particles with a flip and reflects all
spin-$\dn$ particles. From equations
(\ref{tupdif})-({\ref{tpdndif}) we see that the exponents for the
transmission amplitudes are:
\begin{equation}
|t_\up| \sim T^{g_\dn}\,,\qquad
|t_\dn'| \sim  T^{2g_\dn}\,.
\end{equation}
Linearizing equation (\ref{rupup}) for $r_\up'$ in small amplitudes, we have:
\begin{equation}
\frac{ d r_\up'}{d\xi} \approx
\left( g_\dn - g_\perp\right) r_\up' \ \Rightarrow\ r_\up' \sim T^{ g_\dn - g_\perp}\,,
\label{rluplin2}
\end{equation}
which requires $g_\dn > g_\perp$ in order for $r_\up'$ to vanish.
If $g_\dn - g_\perp$ is small, the small quantities neglected in
the right-hand-side of Eq.~(\ref{rluplin2}) will become important.
Therefore, this fixed point holds for $g_\dn - g_\perp$ above some
small quantity. For  larger  $g_\perp$  the system flows to the
spin-flip  reflector fixed point ($ |r_\up'| \rightarrow
\sqrt{v_{F+}/v_{F-}}$).

\section{Discussion and Summary}
\label{final}

Lateral ferromagnetic semiconductor wires with nanoconstrictions
make  it possible to achieve the limit of sharp domain walls
\cite{ruster03}. It has been shown that the constriction itself
does not cause significant reflection of the incident waves
because it only produces a semiclassical potential\cite{khmelnitskii}.
 We may estimate the parameter $\lambda$ of
our model (\ref{model}) by assuming that $\vec{M}(z)= M_0
\cos\theta(z)\hat{\vec{ z}} + M_0 \sin\theta(z){\hat{\vec{ x}}}$
with $\cos\theta(z) = \tanh(z/L)$ \cite{pereira}, where $L$ is the
width of the domain wall.   We then  find
\begin{eqnarray}
\lambda = \frac{JM_0}{\hbar}\int_{-\infty}^{\infty} \sin\theta(z) dz
= \pi\frac{JM_0}{\hbar} L \,,
\label{pi}
\end{eqnarray}
implying that
\begin{eqnarray}
\frac{\lambda}{v_{F+}} = \pi\frac{JM_0}{(\hbar^2 k_{F+}^2/m)} (L
k_{F+}). \label{lam}
\end{eqnarray}
The condition for  the domain wall to be  smaller than the Fermi
wavelength is  $L k_{F+} <2\pi$. The smaller Fermi wavelength is
that of majority spin electrons, $k_{F+}$. On the other hand, for
small   $L k_{F+}$ the barrier becomes a poor spin-flip scatterer.
The ratio  $v_{F-}/v_{F+}$ depends on the polarization degree of
the electron system. We consider now a one-channel system. In a
nonmagnetic system there is a single Fermi momentum $k_F$ for up
and down electrons, and the Fermi energy is
$E_F=\hbar^2k_F^2/(2m)$. Once the system becomes magnetized, the
two new Fermi momenta, $k_{F\pm}$, satisfy  the particle
conservation condition,
\begin{eqnarray}
k_{F+} + k_{F-} = 2k_F\  \Rightarrow\  \frac{k_{F+}}{k_F} +  \frac{k_{F-}}{k_F} =2 \,,
\end{eqnarray}
and the spin-up and spin-down Fermi surfaces must correspond to
the same energy,
\begin{eqnarray}
\frac{\hbar^2 k_{F+}^2}{m} - \frac{\Delta E}{2}
= \frac{\hbar^2 k_{F-}^2}{m} + \frac{\Delta E}{2}\,,
\end{eqnarray}
where $\Delta E/2 = JM_0$ is the Zeeman shift of the bands. From
this we get
\begin{eqnarray}
\frac{k_{F\pm}}{k_F} = 1\pm \frac{\Delta E}{4E_F}\,,
\label{kpminus}
\end{eqnarray}
so that  the ratio $k_{F-}/k_{F+}$ is
\begin{eqnarray}
\frac{k_{F-}}{k_{F+}} =\frac{v_{F-}}{v_{F+}} = \frac{1-(\Delta
E/4E_F)}{ 1+ (\Delta E/4E_F)}\,.
\end{eqnarray}
Inserting (\ref{kpminus})   into Eq.~(\ref{lam}) we obtain
\begin{eqnarray}
\frac{\lambda}{v_{F+}} =  \pi \frac{(\Delta E/4E_F)}{\left[ 1 +
(\Delta E/4E_F)  \right]^2}  \  (L k_+)\,. \label{estima}
\end{eqnarray}
The full polarization limit is  $k_{F-}=0$ and $k_{F+}=2k_F$,
meaning that $\Delta E/4E_F =1$, and then
equation (\ref{estima}) gives
$$
\frac{\lambda}{v_{F+}} \approx 0.79\, L k_{F+}\,.
$$
Typical values for a non fully polarized system are $ E_F = 90$~meV
and  $\Delta E=30$~meV  \cite{ruster03}. In this case we have
$v_{F-}/v_{F+}=0.84$ and  equation (\ref{estima}) gives
$$
\frac{\lambda}{v_{F+}} \approx 0.22\,  L k_{F+} \,.
$$
Therefore, if $L k_{F+}$ is smaller than about $2\pi$, the system
can flow to any of the fixed points described above, especially
the ones described in Section \ref{3conditions}.

The lateral quantization may produce several channels. The higher
channels have larger Fermi wavelength and larger $\Delta E/4E_F$,
so they can be in the spin-flip reflector fixed point. If a
channel of high energy is fully spin polarized, then it
corresponds to $\lambda/v_{F+} = 0.79\, L k_{F+}$. But the
possibility of inter-channel scattering arises. This could be due
to the two following reasons: ({\it i}) electron-electron
interactions (such would require a modification of our theory to
allow for inter-channel scattering); (\it ii}) the impurity
scattering. For the latter to be negligible  we need the electron
mean free path ({\it not} the transport mean free path) to be
larger than the size of the constriction.

In summary, we have studied the effect of electron-electron
interactions on the transmission through a domain wall in a
ferromagnetic wire in the regime in which the wall width is
smaller than the Fermi wavelength. Applying a renormalization
technique to the logarithmically divergent perturbation, we
obtained the scaling equations for the scattering amplitudes. The
$T=0$ fixed points were identified corresponding to: (i) perfectly
insulating wall (with or without complete spin reversal), and (ii)
transparent wall. Both repulsive and attractive interactions were
considered. We have estimated physical parameters for a domain
wall model which may be realized in physical systems. Such
estimates suggest that realistic systems can display the behavior
predicted in the vicinity of the fixed points we have found.

\section*{Acknowledgements}
Discussions with P. A. Lee, A. H. Castro Neto and P. Sacramento
are gratefully acknowledged. M.A.N.A. is grateful to
Funda\c{c}\~ao para a Ci\^encia e Tecnologia for a sabbatical
grant. This research was supported by Portuguese program POCI
under Grant No. POCI/FIS/58746/2004, EU through RTN Spintronics
(contract HPRN-CT-2000-000302), Polish State Committee for
Scientific Research under Grant No. 2~P03B~053~25,  by STCU Grant
No.~3098 in Ukraine and by the German DFG under grant SSP 1165,
BE216/3-2..

\appendix
\section{Generalization of the Wronskian theorem to spinor scattering states}
\label{Wr} The Wronskian theorem\cite{messiah} for (spin
degenerate) scattering states in  one-dimensional systems can be
easily generalized to spinor states in a spin dependent scattering
potential. Let $\psi_1(z)$ and  $\psi_2(z)$ represent  two spinor
scattering states with energies $\epsilon_{1}$ and $\epsilon_{2}$
in the potential $\hat V(z)$. We assume $\hat V(z)$ to be a
$2\times 2$ real matrix,
 as is the case in the Hamiltonian (\ref{model}), and consider a
symmetric mass tensor $\hat m$
(possibly position and spin dependent).
Each spinor satisfies the Schr\"odinger equation
\begin{eqnarray}
\frac{d}{dz} \frac{1}{\hat m}\frac{d\psi_1}{dz}
+ \Big[ \epsilon_1 - \hat V\Big]\psi_1 &=&
\left( \begin{array}{c} 0\\ 0\end{array}\right)\,, \\
\frac{d}{dz} \frac{1}{\hat m}\frac{d\psi_2}{dz}
+ \Big[ \epsilon_2- \hat V\Big]\psi_2 &=&
\left( \begin{array}{c} 0\\ 0\end{array}\right)\,.
\end{eqnarray}
If we multiply the first equation (on the left) by  $\psi_2^t(z)$
 the second equation by  $\psi_1^t (z)$ and subtract the two we obtain
\begin{equation}
\frac{d}{dz}\Big[ \psi_1^t   \cdot\frac{1}{\hat m}\cdot \frac{d\psi_2}{dz}-
\psi_2^t \cdot\frac{1}{\hat m} \cdot  \frac{d\psi_1}{dz} \Big]
= (\epsilon_1 -\epsilon_2) \psi_1^t   \cdot \psi_2\,,
\label{theorem}
\end{equation}
where the dot denotes the matrix (spinor) product ($ \psi_1^t
\cdot \psi_2=\sum_\sigma \psi_{1,\sigma} \psi_{2,\sigma}$). The
expression in square brackets  is a scalar  function of  $z$ and
would be proportional to the Wronskian of the functions $ \psi_1$
and $ \psi_2^*$  in the case where the mass tensor reduces to a
scalar. If the two states  are degenerate ($\epsilon_1=
\epsilon_2$), we conclude from (\ref{theorem})
 that the expression in brackets  is independent of the coordinate $z$:
\begin{equation}
W(\psi_1, \psi_2)\equiv \psi_1^t(z)  \cdot\frac{1}{\hat m} \cdot\frac{d\psi_2}{dz}-
\psi_2^t(z)  \cdot\frac{1}{\hat m}\cdot\frac{d\psi_1}{dz} = const.
\label{wronsk}
\end{equation}
Since the potential matrix $\hat V$ is real, then $\psi_1^*$
(or $\psi_2^*$) also satisfies the Schr\"odinger equation.

The usefulness of the theorem expressed in equation (\ref{wronsk})
is that it allows us to establish general relations between the
scattering amplitudes, independently of the detailed form of the
potential barrier.

If we evaluate (\ref{wronsk}) for  a pair of degenerate scattering
states, say $W(\psi_{k,\up}^*, \psi_{k_-,\dn})$, the result must
be the same for $z<0$ as for $z>0$:
\begin{eqnarray}
W(\psi_{k,\up}^*, \psi_{k_-,\dn})\Big]_{z<0} = W(\psi_{k,\up}^*,
\psi_{k_-,\dn})\Big]_{z>0},
\end{eqnarray}
which yields:
\begin{eqnarray}
&&v_-\ t_\up^*(k)t_\dn'(k_-) + v\ t_\up'^*(k)t_\dn(k_-)\nonumber\\
&+& v\ r_\up^*(k)r_\dn'(k_-) + v_-\ r_\up'^*(k)r_\dn(k_-) = 0.
\label{a6}
\end{eqnarray}
Similarly, calculation of $W(\psi_{k,\up}^*, \psi_{-k_-,\up})$
gives the relation:
\begin{eqnarray}
&&v_-\ t_\up^*(k)r_\dn(k_-) + v\ t_\up'^*(k)r_\dn'(k_-)\nonumber\\ &+&
v\ r_\up^*(k)t_\dn(k_-) + v_-\ r_\up'^*(k)t_\dn'(k_-) = 0 \,,
\label{rel2}
\end{eqnarray}
and  $W(\psi_{k,\up}^*, \psi_{-k_-,\dn})$ gives the relation:
\begin{eqnarray}
&&v_-\ t_\up^*(k)r_\up'(k) + v\ t_\up'^*(k)r_\up(k)\nonumber\\ &+&
v\ r_\up^*(k)t_\up'(k) + v_-\ r_\up'^*(k)t_\up(k) = 0 \,.
\end{eqnarray}
A fourth relation can be obtained from $W(\psi_{k_-,\dn}^*, \psi_{-k_-,\up})$:
\begin{eqnarray}
&&v_-\ t_\dn'^*(k_-)r_\dn(k_-) + v\ t_\dn^*(k_-)r_\dn'(k_-)\nonumber\\ &+&
v\ r_\dn'^*(k_-)t_\dn(k_-) + v_-\ r_\dn^*(k_-)t_\dn'(k_-) = 0\,.
\end{eqnarray}
From  $W(\psi_{k,\up}, \psi_{k_-,\dn})$ we obtain the relation
\begin{eqnarray}
v\ r_\dn'(k_-) = v_-\ r_\up'(k)\,,
\label{a10}
\end{eqnarray}
and $W(\psi_{k,\up}, \psi_{-k_-,\up})$ gives
\begin{eqnarray}
v\ t_\dn(k_-) = v_-\ t_\up(k)\,.
\end{eqnarray}

Considering  a state  and its conjugate, $W( \psi_{k,\sigma}^*,
\psi_{k,\sigma})$ gives the conservation of the charge current,
\begin{equation}
v\ =\  v_- |t_\up|^2 +\  v |t_\up'|^2 +\  v |r_\up|^2 +\  v_- |r_\up'|^2\,,
\label{conserv}
\end{equation}
for $\sigma=\up$, and
\begin{equation}
v_-\ =\  v |t_\dn|^2 +\  v_- |t_\dn'|^2 +\  v_- |r_\dn|^2 +\  v |r_\dn'|^2\,,
\end{equation}
for $\sigma=\dn$.

\section{Formulation for spin dependent electron effective masses}
\label{masses}
Electron interactions such as $g_4$ and  $g_2$,
which describe forward  scattering between particles moving in the
same direction, may produce  renormalization of the electron's
effective mass.\cite{Solyom} The latter  could depend on spin
orientation because the  Fermi surfaces of spin-up and spin-down
electrons are different. These effects can be taken into account
from the beginning by rewriting  the Hamiltonian (\ref{model}) in
a more general form,
\begin{equation}
\hat H_0 = -\frac{\hbar^2}{2}\frac{d}{d z}
\frac{1}{\hat m(z)} \frac{d}{d z} + \hbar V \delta(z)
-J M_z(z)\hat \sigma_z - \hbar \lambda   \delta(z)\hat \sigma_x\,,
\end{equation}
where, in the kinetic energy  term, we allow for a position and spin
dependent effective mass tensor, $\hat m(z)$.
The tensor may take the form:
\begin{equation}
\hat m(z) = \left( \begin{array}{cc} m_\up(z) & 0 \\
0 & m_\dn(z) \end{array} \right) \,,
\end{equation}
with
$$
m_\up(z) = m_+ \Theta(-z) + m_-\Theta(z)
$$ and
$$
 m_\dn(z) = m_- \Theta(-z) + m_+\Theta(z)\,,
$$
where $\Theta(z)$ denotes the Heaviside function.

The appropriate mass values must be used in Eqs.~(\ref{dois})-(\ref{kmn}).
The expressions for the scattering
eigenstates and transmission amplitudes given in the main text
remain unchanged if we take into account that  the velocities must
be calculated considering the renormalized masses.




\end{document}